\documentclass[times,twocolumn,final,nopreprintline]{elsarticle}

\usepackage{framed,multirow}

\usepackage{amssymb}
\usepackage{latexsym}

\usepackage{url}
\usepackage{xcolor}
\usepackage{amsmath}
\usepackage{array} 
\usepackage{tabularx}
\usepackage{ragged2e}
\usepackage{booktabs}
\usepackage{rotating}
\usepackage{float}
\usepackage{lscape}
\usepackage{pdflscape}
\usepackage{supertabular}
\usepackage{longtable}
\usepackage{changepage}
\usepackage[hidelinks]{hyperref} 
\usepackage{hyperref}
\usepackage{soul}
\usepackage{stfloats}
\hypersetup{
  colorlinks   = true, 
  urlcolor     = blue, 
  linkcolor    = blue, 
  citecolor   = red 
}
\definecolor{newcolor}{rgb}{.8,.349,.1}

\newcommand{\etal}{\textit{et al. }}
\begin{document}

\begin{frontmatter}

\title{\textit{CellOMaps}: A Compact Representation for Robust Classification of Lung Adenocarcinoma Growth Patterns}

\author[a1]{Arwa Al-Rubaian\corref{cor1}}
\cortext[cor1]{Corresponding author:}
\ead{arwa.alrubaian@warwick.ac.uk}
\author[a1]{Gozde N. Gunesli}
\author[a2]{Wajd A. Althakfi}
\author[a3]{Ayesha Azam}
\author[a1,a3,a4]{David Snead}
\author[a1,a4]{Nasir M. Rajpoot}
\author[a1]{Shan E Ahmed  Raza}
\ead{shan.raza@warwick.ac.uk}

\address[a1]{Tissue Image Analytics Centre, Department of Computer Science, University of Warwick, UK.}
\address[a2]{Histopathology unit, department of Pathology, king Saud University, Riyadh,Kingdom of Saudi Arabia.}
\address[a3]{Department of Histopathology, University Hospitals Coventry and Warwickshire NHS Trust, Coventry, UK.}     
\address[a4]{Histofy Ltd, Coventry, UK.}


\begin{abstract}
Lung adenocarcinoma (LUAD) is a morphologically heterogeneous disease, characterized by five primary histological growth patterns. The classification of such patterns is crucial due to their direct relation to prognosis but the high subjectivity and observer variability pose a major challenge. Although several studies have developed machine learning methods for growth pattern classification, they either only report the predominant pattern per slide or lack proper evaluation.
We propose a generalizable machine learning pipeline capable of classifying lung tissue into one of the five patterns or as non-tumor. The proposed pipeline's strength lies in a novel compact Cell Organization Maps (cellOMaps) representation that captures the cellular spatial patterns from Hematoxylin and Eosin (H\&E) whole slide images (WSIs). The proposed pipeline provides state-of-the-art performance on LUAD growth pattern classification when evaluated on both internal unseen slides and external datasets, significantly outperforming the current approaches. In addition, our preliminary results show that the model’s outputs can be used to predict patients Tumor Mutational Burden (TMB) levels.

\end{abstract}


\end{frontmatter}

\newcolumntype{P}[1]{>{\RaggedRight\footnotesize}p{#1}}
\section{Introduction}
\label{sec:intro}
\begin{figure*}
\centering
\includegraphics[width=\textwidth]{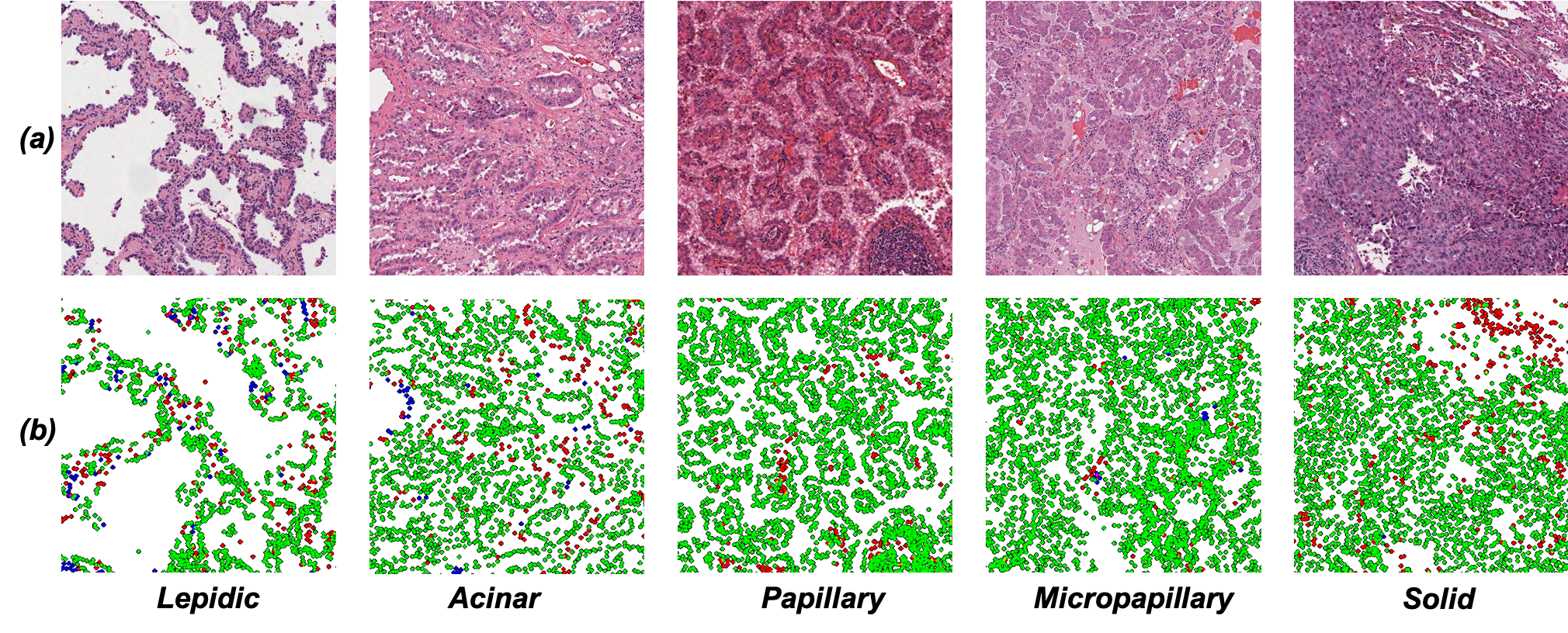}
\caption{Sample images of lung adenocarcinoma (LUAD)  growth patterns: (a) the H\&E image, (b) the corresponding \textit{CellOMaps}, where green, red, and blue dots denote neoplastic cells, connective cells, and non-neoplastic cells, respectively.}
\label{patternsVisual}
\end{figure*}

Lung cancer is one of the most prevalent forms of cancer worldwide, being the second most common cancer (after breast cancer) and the leading cause of cancer-related deaths, accounting for approximately 18\% of all cancer deaths globally \cite{cancerStats}. Around 85\% of all reported lung cancer cases are classified as non-small cell lung cancer (NSCLC), with Lung Adenocarcinoma (LUAD) as its most prevalent subtype. According to the latest World Health Organization (WHO) classification, invasive non-mucinous LUAD grows into five primary histological growth patterns: lepidic, acinar, papillary, micropapillary, and solid \cite{whoClassification}. Examples of the five growth patterns are shown in Figure \ref{patternsVisual}(a). These patterns are defined and can be differentiated by the arrangement of tumor cells. In the lepidic pattern, tumor cells line the alveolar walls preserving the original structure of the lung. The tumor cells in the acinar pattern form glandular shapes, usually separated by stroma. While in the papillary pattern, tumor cells grow into small finger like projections, called papillae, having a fibervascular core. Whereas, the micropapillary pattern is defined by small clusters of tumor cells with no fibervascular core. The solid pattern appears as sheets of unstructured tumor cells \cite{whoClassification}.

The identification of such patterns in a LUAD tumor is crucial as these patterns carry prognostic value and affect the patient outcomes, with lepidic having the most favorable prognosis, followed by acinar and papillary, whereas solid and micropapillary have been known to have the worst prognosis of all \cite{grading}.
Despite the clear definition of these growth patterns, their visual identification and classification in a tumor tissue is very subjective and has high  inter- and intra-observer variability due to their overlapping nature and lack of quantitative measures \cite{observerVariability}. The solid pattern is relatively easy to identify and has high inter-observer agreement, whereas papillary and micropapillary patterns suffer from the lowest inter-observer agreement \cite{observerVariability}.

As cancer is heterogenous, within the same tumor, LUAD often presents a varied combination of multiple growth patterns. The WHO recommends a three-tiered grading system for lung adenocarcinomas, which involves assessing the predominant histological growth pattern and determining the presence or absence of high-grade patterns (solid and micropapillary). This approach has been shown to provide more accurate information about patient outcomes \cite{whoClassification}. However, this can be challenging, as a small proportion of high grade patterns can easily be missed during slide examination.  
Additionally, this diagnostic method fails to consider the diverse patterns and their spatial arrangement within the tumor, which could provide more accuracy in diagnosis and improve prognostic predictions for patients. The automation of growth pattern classification using machine learning would be a valuable addition to pathology, enhancing the precision and objectivity of the task while alleviating its labor-intensive and time-consuming nature.

In this paper, we aim to address the aforementioned challenges by developing a machine learning pipeline that can accurately distinguish between the five different growth patterns and the normal tissue with an average overall accuracy of 0.81. Moreover, careful qualitative analysis of the results show that the majority of mis-classification was due to the presence of mixed patterns in those tile. This is achieved by introducing a novel image representation that we term as the \textit{Cell Organization Maps} (or \textit{CellOMaps} for short), which is a transformation and compression of the Hematoxylin and Eosin stained (H\&E) whole slide image (WSI). Through extensive and appropriate evaluation, we demonstrate that the proposed \textit{CellOMaps} preserve the sufficient amount of information for the task of growth pattern classification, removing the noisy and irrelevant details of the H\&E image. 

\textit{CellOMaps} effectively capture the cellular organization of the tissue at a sufficient level of detail. It efficiently preserves nuclei locations and types, where image channels correspond to different nuclei types and each nucleus is represented by its centroid. This representation allows deep learning models to focus on how cells spatially interact within their microenvironment, by minimizing the noise and complex details of an H\&E image.

The main contributions of this work are as follows:
\begin{enumerate}

    \item We introduce \textit{CellOMaps}, a novel compressed WSI representation (approximately 10:1 compression), that captures and emphasize the cellular arrangement, allowing deep learning models to focus on relevant details that might be challenging to extract directly from H\&E during training for pathology-specific tasks. To the best of our knowledge, ours is the first study that uses such representation for tissue classification. 

    \item We evaluate \textit{CellOMaps} on three external datasets that have been annotated by pathologists who did not label any of the training samples. Due to the fact that \textit{CellOMaps} capture only necessary information and ignore unnecessary details, they show good generalizability.

    \item  We show that our proposed pipeline achieves the state-of-the-art performance when properly evaluated using the patient-level splitting approach. With an average overall acuraccy of 0.81, the pipeline outperforms baselines and current growth pattern classifiers with a significant margin.    

    \item We present indicative results demonstrating how the model's outputs can effectively stratify patients according to their Tumor Mutational Burden (TMB) levels. 

\end{enumerate}

\section{Related Work}
\label{sec2}
The invention of digital scanners have revolutionized the field of pathology, introducing digital pathology. The digitized WSIs along with the advancement in machine learning algorithms enabled the development of various segmentation and classification algorithms that can analyze and extract information from such images. Researchers have investigated the use of machine learning to perform various pathology image analysis tasks ranging from tumor segmentation and subtype classification \cite{coudray2018classification}, to mutation prediction \cite{neda2024social} and the classification of growth patterns in a LUAD tissue slide stained with H\&E \cite{alsubaie_growth_2023}\cite{Rina_tailoritingPretext}\cite{darmouth}.

The classification of LUAD growth patterns is an active area of research. The majority of machine learning algorithms designed for this task predict a single pattern for each WSI, namely the predominant pattern. 

Wei \etal \cite{cao2023e2efp} used the Dartmouth-Hitchcock Medical Center dataset to train a ResNet-18 to classify 224×224 patches into one of the growth patterns. For each WSI, the majority class was reported as the dominant pattern, along with patterns spanning more than 5\% of the slide as minor patterns. They report an F1-score of 90.4\% for their dominant pattern prediction \cite{darmouth}. Cao \etal extracted features from intermediate layers of a deeper ResNet-34, they performed feature reduction and aggregation to obtain patch level feature representations. A sliding kernel with an underlying max operation was then used to aggregate patch level prediction, followed by a fully connected layer to produce a single class for each WSI \cite{cao2023e2efp}.

These methods achieve good performance and can be used to assist pathologists in their routine work. However, they do not take into account the heterogeneous composition of growth patterns and their spatial relationships, which can add more precision to the diagnosis and enhance the prediction of patient outcome. Predicting the different growth patterns in a WSI can be done either by pixel classification (semantic segmentation) or by super-imposing the results of a tile-based classifier on the WSI. A few attempts have explored using semantic segmentation for defining the different growth patterns, ranging from using ensemble U-Net \cite{shao2023pixel} to exploiting multi magnifications \cite{campanella2022h}. Using average filters of different sizes, Pan \etal \cite{tracerX2022cross} produced three images of varying detail. Each image was fed into a segmentation stream and results from each stream were used to guide the segmentation of the more detailed  stream via attention \cite{tracerX2022cross}.

Semantic segmentation is often used to detect the boundaries of objects or components in an image and has been successful in several pathology applications such as tissue semantic segmentation \cite{tissueSeg} and nuclei segmentation \cite{hoverNet}. However, growth patterns often do not have clear boundaries or outlines, as usually they smoothly transition from one pattern to another with borderline areas in between. Thus projecting tile classification results on WSIs would be more suitable for such application. Early works in this area designed models to predict the presence or absence of a single pattern \cite{microp_detect} or classify a tile into a selected subset of the growth patterns, with the exclusion of the challenging papillary and lepidic patterns  \cite{fourPatterns}. 

Only a limited number of studies have developed tile-based classifiers capable of predicting all five growth patterns. Alsubaie \etal \cite{alsubaie_growth_2023} reported that using views from two magnifications perform better than a single magnification. The authors stacked the images from two different magnification namely 10× and 20×, and aligned them in the center, then fed the resulting 224×224×6 patches to a modified ResNet-50 tile classifier. The model achieved a mean accuracy of 91\% using data splits obtained at the tile-level \cite{alsubaie_growth_2023}.

Ding \etal \cite{Rina_tailoritingPretext} also used a ResNet-18 backbone, but pre-trained three networks each with a pathology-specific pre-text task \cite{Rina_tailoritingPretext}. The first task was to predict if the higher magnification tile of a 6-channel image is in the first or last channels and if it belongs to the lower magnification tile. The second task was given a 6-channel image composed of a tile and a crop of that tile, the model predicts the grid location the crop is from. The third task was predicting the Eosin stain from a Hematoxylin stained image. For the downstream task of classifying a tile as normal or assigning it a growth pattern, these three models were finetuned and an ensemble network was created, where the final tile classification result was a weighted average of the probabilities from all three models. The authors trained and tested their model on tiles from TCGA, NLST, and used tiles from CPTAC as external testing. They report F1-score ranging from 0.85 to 0.92 on the internal testset, but the performance degrades dramatically on the external testing set \cite{Rina_tailoritingPretext}.

The aforementioned tile classification studies adopt a tile-based data splitting approach. Consequently, a tile in the test set might originate from the same WSI or even be adjacent to a tile used in training the model. This notably enhances the classifier’s performance because tiles from the same pattern within a WSI possess substantial visual resemblance, unlike similar pattern tiles from other WSIs. As expected, these models often encounter failure when new data or WSIs that have not been seen before are used for external validation. 

The only study we found in the literature adopting a patient-level splitting is Sadhwani \etal \cite{sadhwani2021comparative}, who used Inception-V3 to classify an 8×8 mid-region using tiles of size 512×512 at 10× magnification into nine histological subtypes (including five growth patterns).These classification results were then used as part of a feature vector input for a TMB classification model. The proposed approach performed well on predicting growth patterns with a mean tile-level AUC-ROC of 0.9\%. however, the class imbalance problem was not considered while assessing their model's performance. Due to the high class imbalance in such patterns, AUC-ROC may not be the best choice for assessment \cite{sadhwani2021comparative}.

\section{The Proposed Method}
\label{sec3}
\begin{figure*}[p]  
\centering
\includegraphics[width=\textwidth, height=0.85\textheight,keepaspectratio]{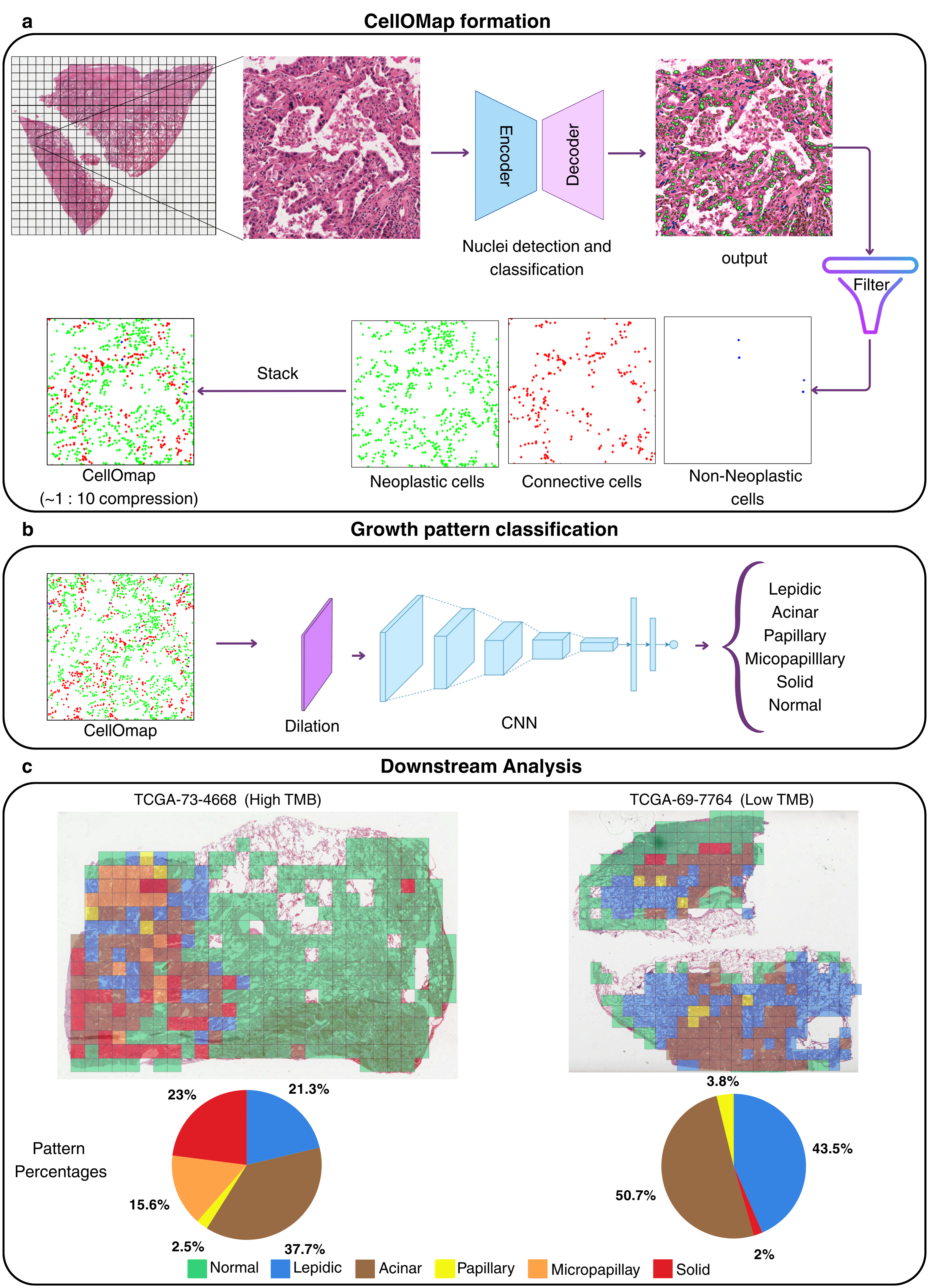}

\caption{An overview of the proposed model for growth pattern classification. (a) The formation of the CellOMaps: First, nuclei in a slide are detected and classified using HoVer-Net. Then relevant nuclei are filtered and stacked (each class in a channel) to form a 3-channel image; (b) The CellOMaps is input to a convolution neural network (ResNet-50) with an extra dilation layer, for growth pattern classification; (c) Projection of the predicted patterns on sample WSIs, give an indication that growth patterns provide insight into TMB levels, distinguishing between high and low TMB.}
\label{model}
\end{figure*}

Inspired by the WHO definition of the five LUAD growth patterns \cite{whoClassification}, which is primarily dependent on the arrangement of neoplastic cells in the tissue and in some patterns their placement  with the connective tissue, we develop a  novel pipeline for LUAD growth pattern classification. An overview of the proposed pipeline is shown in Figure \ref{model}. It comprises two major parts: the construction of the \textit{CellOMaps}, followed by the prediction of the different growth patterns, which are explained in detailed in the following subsections.

\subsection{CellOMaps construction}
Our model relies on the simplified and compact representation of the WSI, referred to as \textit{CellOMaps}. To obtain such representation first, the nuclei need to be identified and classified. Then the nuclei masks are processed to form the final \textit{CellOMaps}.

As illustrated in Figure \ref{model} (a), we employed HoVer-Net \cite{hoverNet} to find the location and type of nuclei cells in a given WSI. HoVer-Net was trained on the PanNuke dataset \cite{pannuke} composed of a large amount of histology images obtained from 19 different tumor sites, including lung. The entire WSI was processed at 0.5 $\mu$m per pixel (mpp), roughly equivalent to 20× magnification.

Then we filtered the nuclei classified by HoVer-Net to include only the neoplastic epithelial, non-neoplastic epithelial, and connective cell nuclei. This selection was based on the definition of such growth patterns \cite{bookpathology}, where the arrangement of neoplastic epithelial cells is the key visual aspect differentiating the growth patterns apart. The non-neoplastic epithelial cells were included as they define the normal tissue. The connective cells were incorporated to aid in the classification of the acinar pattern, but more importantly the papillary pattern, which is defined as tumor papillea arranged around a fibrovascular core.
Nuclei coordinates were scaled to 2 mpp (equivalent to 5× magnification). A binary mask was then created for each nuclei class, where a value of one was assigned to pixels corresponding to nucleus centroids. 
To form the final 3-channel \textit{CellOMaps}, the three binary masks were normalized to the RGB space (with values ranging from 0 to 255) and stacked, with each mask representing a separate channel. As illustrated in Figure \ref{patternsVisual} (b), the visual distinction between the different growth patterns is maintained in the generated \textit{CellOMaps}. Figure \ref{model} (a) shows the formation of \textit{CellOMaps} at the tile level for demonstration purpose. However the generation of maps is performed at the WSI level in the original pipeline. 

\subsection{Growth pattern classification}
We fine-tuned a ResNet-50 model, pretrained on ImageNet, by modifying the final fully connected layer to classify images into one of six classes: lepidic, acinar, papillary, micropapillary, solid, and normal. To improve centroid visibility in the images, we introduced a dilation layer as the initial layer of the network. This dilation layer consists of a 5×5 kernel with all ones, followed by a max pooling operation. The network was fed with \textit{CellOMaps} tiles of size 448×448. Data augmentation was performed during training by applying random horizontal and vertical flips. We used focal loss \cite{focalLoss} along with Adam optimizer and a learning rate of  10\textsuperscript{-5} to train the model.  
Focal loss was used to address the class imbalance issue, by assigning higher weights for the loss of hard to classify (minority) classes, so that the gradient is not dominated by easy classes with many correctly classified samples. Focal Loss modifies the cross-entropy loss by adding a modulating factor as follows,
\begin{equation}
\text{FL}(p_t) = -\sum_{i=1}^K \alpha_i (1 - p_i)^\gamma y_i \log(p_i)
\end{equation}
where $p_i$ represents the predicted probability for the class $i$,
\begin{equation}
y_i =
\begin{cases}
  1 & \text{if } y = i \\
  0 & \text{otherwise}
\end{cases}
\end{equation}
and the modulating factor $(1 - p_i)^\gamma$ is introduced to reduce the loss assigned to well-classified examples ($p_i$ close to 1) and place more emphasis on hard examples ($p_i$ close to 0). The hyper-parameter $\gamma$ controls the strength of this modulation. When $\gamma = 0$, Focal Loss simplifies to the standard cross-entropy loss. As $\gamma$ increases, the effect of the modulating factor becomes stronger, focusing the model more on misclassified or difficult examples. Finally, $\alpha_i$ is an optional weighting factor that can be applied to balance the loss between classes.

\section{Experimental Results}
\label{sec4}
\subsection{Datasets}
\label{subsec41}
For this study, we utilized four datasets: TCGA-LUAD, NLST, UHCW, and the ANORAK training data from the TRACERx project. Detailed descriptions of these datasets and their respective annotation processes are provided in the subsequent subsections. Only cases of non-mucinous lung adenocarcinoma were included from each dataset. A summary of the datasets and their class distributions is presented in Table \ref{tab:datasets}.

\begin{table*}
\caption{Total Number of WSIs and tiles per class for each dataset. }
    \centering
    \vspace{0.5em}
    \resizebox{\textwidth}{!}{%
    \begin{tabular}{@{}lcccccccc@{}}
    \hline
       Dataset  & WSIs & Solid & Acinar & Papillary & Micropapillay & Lepidic & Normal & Total tiles \\ \hline
       TCGA-LUAD  & 46 & 277 & 310 & 109 & 133 & 67 & 396 & 1292\\
       NLST  & 78 & 133 & 122 & 111 & 49 & 123 & 152 & 690\\
       UHCW  & 19 & 156 & 168 & 323 & 77 & 68 & 395 & 1187\\
       ANORAK data & - & 132 & 108 & 33 & 7 & 84 & 127 & 491 \\
       \hline
    \end{tabular}
    }
    
    \label{tab:datasets}
\end{table*}

\subsubsection{TCGA-LUAD:}
\label{secTCGA}
 A total of 46 diagnostic WSIs obtained from 17 different centers were randomly selected and downloaded from the publicly available Cancer Genome Atlas \cite{TCGA}.  All slides were annotated by three expert pulmonary pathologists, and ground truth labels were acquired by the conscience of at least two pathologists. The annotating process comprised three main steps: First two pathologists were separately asked to annotate regions of the WSIs best expressing the different growth patterns. The only requirement was to identify regions as large as possible, which are relatively pure (i.e, containing a single pattern). Then we identified non-overlapping areas between individual annotators. We requested both the pathologist to provide an annotation for those non-overlapping regions independently. A third pathologist was asked to provide a pattern for all regions where the first two pathologist disagreed. Finally, the concession of at least two annotators was used as ground truth labels. To maintain consistency with the other datasets, the number of tiles reported in Table \ref{tab:datasets} corresponds to tiles of size 1024×1024 at 0.5 mpp. However it is worth noting that \textit{CellOMaps} allow us to fit a wider context into memory and thus the number of patches vary for this dataset based on the tile size used.

\subsubsection{UHCW: }
\label{secUHCW}
A local dataset obtained from the University Hospital Coventry and Warwickshire (UHCW) comprising of 1,187 tiles acquired from 19 WSIs, each from a different patient. This dataset was annotated following the same process used to annotate the TCGA-LUAD slides mentioned previously in Section \ref{secTCGA} above. All tiles have a size of 1024×1024 at 20× magnification.

\subsubsection{NLST:}
\label{secNLST}
A total of 690 tiles extracted from 78 slides from the publicly available National Lung Screening Trial (NLST) dataset \cite{NLST}. This dataset was constructed and annotated by pathologists from University of California, Los Angeles \cite{Rina_tailoritingPretext}. All tiles have a size of 1024×1024 at 20× magnification. 

\subsubsection{ANORAK training data:}
\label{secTracerX}
This dataset is a subset of the training dataset used to train the segmentation model ANORAK \cite{ANORAK}. It was collected and annotated by the TRACERx (TRAcking Cancer Evolution through therapy (Rx)) project \cite{tracerx}. This dataset was cleaned and curated to fit the classification problem. First all tiles were scaled to 0.5 mpp. Tiles expressing more than one pattern, and tiles with major artifacts such as blurring or no tissue were removed. Finally, we unified the tile sizes to 1024×1024, where images with insufficient tissue amount (less than 1000×1000 of tissue) were removed. For extremely large tiles, a maximum of two non-overlapping tiles were generated. The final cleaned dataset contained 491 tiles. \par

A quality check was conducted on the HoVer-Net output on each of our datasets separately. In both the TCGA-LUAD (refer to section \ref{secTCGA}), NLST dataset (refer to section \ref{secNLST}), and ANORAK training data (refer to section \ref{secTracerX}) the number of misclassification was minimal and did not affect the overall structure of the patterns. However, in the UHCW dataset (refer to section \ref{secUHCW}), we noticed that a great majority of the neoplastic cells were classified as necrosis in a couple of slides where there was extreme difference in stain appearance. Further investigation showed that around 90\% of the cells classified as necrosis were in fact neoplastic cells, with the remaining minority to be inflammatory cells. This observation was verified by our pathologist and thus all cells classified as necrosis were treated as neoplastic cells in this dataset.

\subsection{Comparative evaluation}
\label{subsec43}
The performance of the proposed model was evaluated through internal cross validation and external testing. For both experiments, we compare the performance of our proposed \textit{CellOMaps} approach with five state-of-the-art and baseline models including: 
\begin{enumerate}
    \item \textbf{A multi-resolution approach:} where we proposed to train two identical ResNet-34 models: one on 1024×1024 H\&E tiles at 20× magnification and the other on 1024×1024 H\&E tiles at 5× magnification. The final classification result is determined by selecting the prediction of the model with the highest confidence when fed with tiles having the same center. This approach is similar to the way we observed pathologists typically analyze these patterns, where they make an initial decision at lower magnifications and then confirm or alter their decision at higher magnifications.
    \item \textbf{ResNet-50:} The widely used baseline \cite{darmouth}\cite{alsubaie_growth_2023}\cite{ANORAK}\cite{deepluad}\cite{resNetAllNet}, in which we apply transfer learning from the pre-trained ImageNet weights. The input to the network is 1024×1024 H\&E tiles at 20× magnification.
    \item \textbf{AlSubaie \etal \cite{alsubaie_growth_2023}:} The model is  a modified ResNet-50 that takes as an input a 6-channel image. The image is formed by stacking a tile at 20× magnification, and at 10× magnification of an H\&E image and aligning them in the center .
    \item\textbf{Ding \etal \cite{Rina_tailoritingPretext}:} The model consists of three ResNet-18 networks each pretrained on a different pathology-specific task. It takes as an input a 1024×1024 tile at 20× magnification, passes it through the pretrained models separately, then uses weighted averaging to ensemble the final prediction.  
    \item \textbf{ViT-16:} We use the Hugging Face implementation of the vision transformer \cite{ViT} to classify 448×448 tiles extracted at 5× magnification.
   
\end{enumerate}

\begin{figure*}
\centering
\includegraphics[width=0.6\textwidth,keepaspectratio]{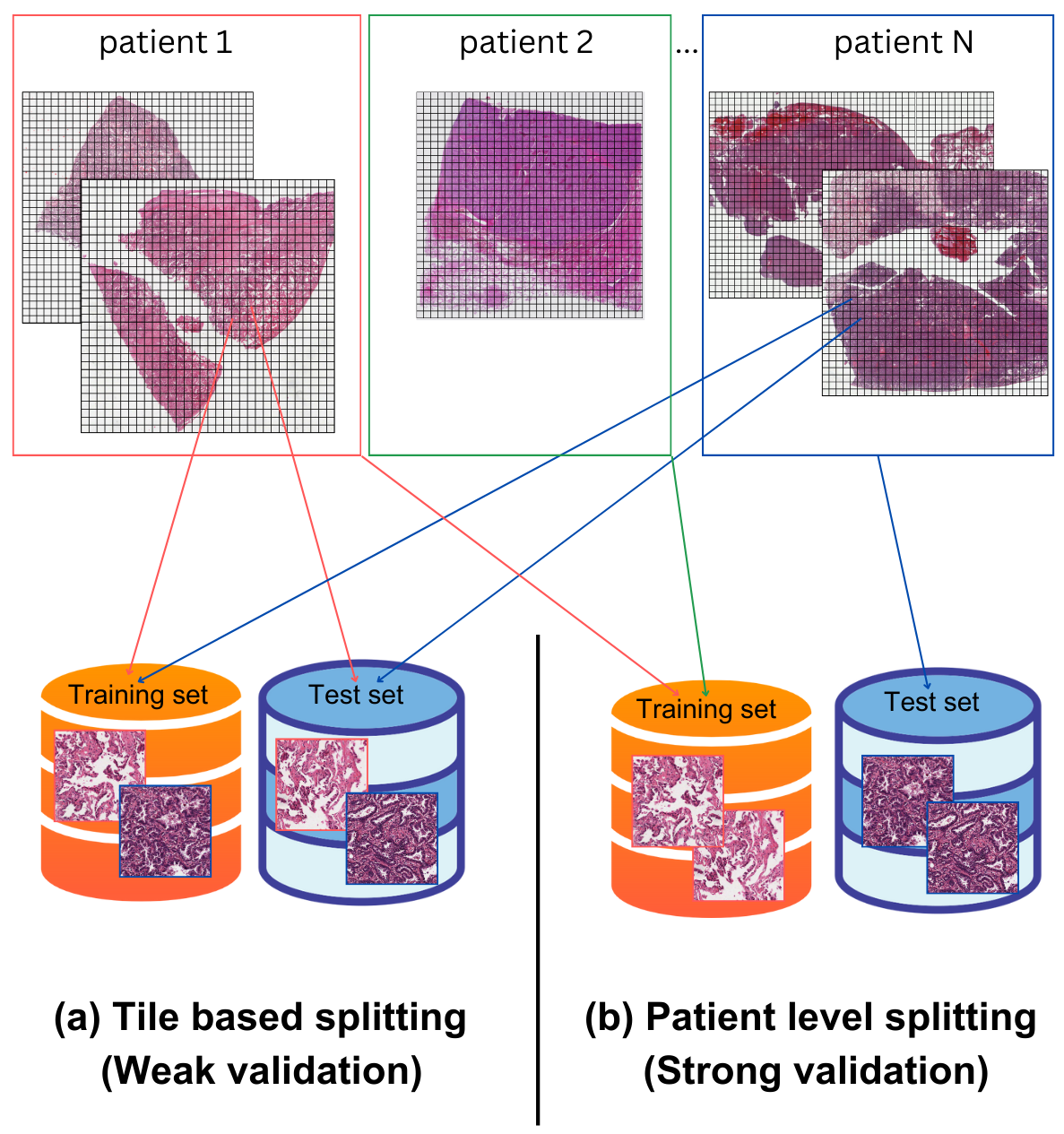}
\caption{The difference between tile-based splitting used widely in the literature (a) and the appropriate patient level splitting employed in this study (b). In (a) tiles from a single patient could be included in both the training and test set. In (b), there is a strict boundary between training and test sets and mixing of tiles is not allowed.}
\label{weakstrong_validation}
\end{figure*}

\subsubsection{Patient-level Cross Validation}

The validation approach employed by growth pattern classifiers in the current literature adopt a tile-level splitting approach (illustrated in Figure \ref{weakstrong_validation} (a)), to create their train/test splits, where tiles from all WSIs in the dataset are mixed then divided into training and testing sets. This method does not ensure a robust evaluation of the model, as visually similar, adjacent tiles from the same WSI may end up in different partitions (training or testing). Consequently, the high performance metrics reported using this validation approach often indicate over fitting.

To provide a strong and more robust evaluation we adopt a patient-level splitting approach on the TCGA-LUAD dataset (illustrated in Figure \ref{weakstrong_validation} (b)), where splitting of data is done at the patient level, before tiling. We consider this to be a stronger evaluation, where we ensure that the tiles in the test set originate from WSIs of patients that none of their WSIs (or part of it) have been utilized for training, thus minimizing leakage and creating a \textbf{truly unseen} test set.

We trained and evaluated each model 5 times. In each trial approximately 10 patients were randomly sampled from the dataset as the unseen test set. Stratified sampling was used to ensure the presence of all patterns in the unseen test set. The remaining slides were then tiled and split into 80\% training and 20\% validation. 

\begin{figure}
\centering
\includegraphics[width=0.5\textwidth,keepaspectratio]{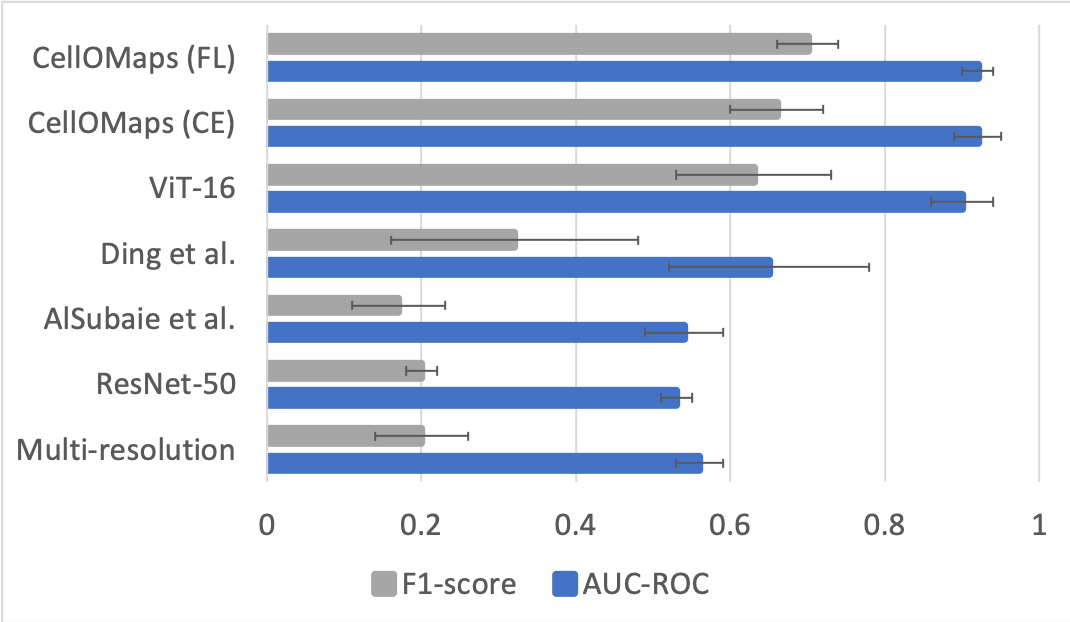}
\caption{Average and standard deviation of AUC-ROC and macro F1-scores for growth pattern classification on TCGA-LUAD using cross validation on the patient level. (CE): cross entropy, (FL): Focal loss}
\label{internal_validation}
\end{figure}

Figure \ref{internal_validation} summarizes the results of the internal validation on the TCGA-LUAD cohort. It is clearly evident that the proposed \textit{CellOMaps} approach outperforms all other approaches proposed in the literature by approximately 50\% in F1 score, achieving  an average AUC-ROC of 0.92 and an average F1-macro score of 0.7. Even when adopting the widely used cross entropy loss function, \textit{CellOMaps} outperforms all other methods. 
The simplification of the input from complex H\&E images to images containing only the nuclei type and location information, caused the model's learning to focus on such information without getting diverted to other non-relevant aspects including stain and tissue related features.  

Table \ref{tab:f1_internal} reports the average F1-scores and standard deviations detailed for each growth pattern. It can be seen that the proposed model predicts all patterns with very high precision. The major confusion is between the papillary and micropapillary patterns, which have high disagreement levels even between trained pathologists.

\begin{table*}
    \caption{Average and standard deviation of per class accuracy for growth pattern classification on TCGA-LUAD using cross validation on the patient level; (CE): cross entropy, (FL): Focal loss.}
    \centering
    \vspace{0.5em}
    \resizebox{\textwidth}{!}{%
    \begin{tabular}{@{}lcccccccc@{}}
    \hline
        Method & solid & Acinar & Papillary & Micropapilaary & lepidic & Normal & Overall \\ \hline
         Multi-resolution &0.45 ± 0.19&0.36 ± 0.09& 0.12 ± 0.12&0.05 ± 0.04&0.02 ± 0.03& 0.38 ± 0.22 & 0.27 ± 0.07\\
         ResNet-50 &0.35 ± 0.14&0.45 ± 0.24&0.10 ± 0.18&0.10 ± 0.14&0.02 ± 0.03&0.42 ± 0.16 & 0.26 ± 0.04\\
         AlSubaie \etal \cite{alsubaie_growth_2023}&0.23 ± 0.03&0.43 ± 0.13&0.0 ± 0.0&0.18 ± 0.32&0.05 ± 0.10&0.52 ± 0.24 & 0.27 ± 0.12\\
         Ding \etal \cite{Rina_tailoritingPretext}&0.59 ± 0.22&0.32 ± 0.16&0.49 ± 0.39&0.34 ± 0.30& 0.09 ± 0.10&0.22 ± 0.12 & 0.40 ± 0.20\\
         ViT-16 \cite{ViT} & \textbf{0.94  ± 0.04}&0.73 ± 0.17 &0.30 ± 0.23&0.41 ± 0.17&0.60 ± 0.24&\textbf{0.99 ± 0.01} & 0.78 ± 0.08\\
         \hline
         \textit{CellOMaps} (CE)&\textbf{0.90 ± 0.07}&\textbf{0.82 ± 0.08}&0.32 ± 0.20&0.43 ± 0.18&0.62 ± 0.15&0.96 ± 0.05 & 0.76 ± 0.03 \\  
         \textbf{\textit{CellOMaps} (FL)} &0.87 ± 0.13 &0.74 ± 0.11&\textbf{0.45 ± 0.2}&\textbf{0.51 ± 0.13}&\textbf{0.77 ± 0.09}&\textbf{0.97 ± 0.01} & \textbf{0.81 ± 0.02} \\ \hline
    \end{tabular}
    }   
    \label{tab:f1_internal}
\end{table*}

\subsubsection{External Validation}

For external validation, we tested the model on three datasets obtained from different sources and annotated by different pathologists, the NLST dataset, the UHCW cohort, and the ANORAK training data. In all experiments the methods were trained on the entire TCGA-LUAD and tested on the target external dataset. Due to the annotation format of the NLST dataset and the limited annotated area of the UHCW dataset we use smaller \textit{CellOMaps} tiles of 256×256 at 2 mpp. 

The external validation results on the UHCW, and NLST datasets are detailed in Table \ref{tab:UHCW_extVal} and \ref{tab:NLST_extVal} respectively, comparing our model to the state-of-the art methods and baselines. 
The highest F1-score on the UHCW dataset was 0.57 achieved by the proposed \textit{CellOMaps} outperforming the second best model (ViT-16) by 7\%. All other methods fall behind by 20\% or more. However, on the NLST dataset ViT slightly outperforms \textit{CellOMaps}. That can be credited to its slightly better performance in classifying normal, solid, and acinar tiles, which represent a great portion of dataset. However for the harder classes, with fewer examples such as micropapillary and papillary \textit{CellOMaps} provides higher precision.  These results gives an indication that our proposed method generalizes well compared to the other approaches, given its performance on the external validation sets where the data came from different hospitals and was annotated by different pathologists. 

We experience similar behavior when evaluating the models on the external ANORAK dataset. The proposed \textit{CellOMaps} approach achieved a macro F1-score of 0.61 and an accuracy of 0.78. Where the major misclassification was in the papillary samples. we provide comparative results in Table \ref{tab:ANORAK_extVal}. Due to the unavailability of the original slides we exclude models that require views from different magnifications. 

\begin{table}
    \caption{External validation results for growth pattern classification on the UHCW cohort.}
    \centering
    \vspace{0.5em}
    \begin{tabular}{lccc}
    \hline
       Method  & AUC-ROC & Accuracy & F1-score \\ \hline
       Multi-resolution & 0.75 & 0.48 & 0.38 \\
       ResNet-50 & 0.64 & 0.30 & 0.22 \\
       AlSubaie \etal \cite{alsubaie_growth_2023}  & 0.67 & 0.47 & 0.30 \\
       Ding \etal \cite{Rina_tailoritingPretext}  & 0.60 & 0.33 & 0.08 \\
       ViT-16 \cite{ViT} & 0.73 & 0.61 & 0.50  \\ \hline
       \textbf{\textit{CellOMaps}}  & \textbf{0.89} & \textbf{0.66} & \textbf{0.57} \\ \hline
    \end{tabular}
    \label{tab:UHCW_extVal}
\end{table}

\begin{table}
    \caption{External validation results for growth pattern classification on the NLST dataset.}
    \centering
    \vspace{0.5em}
    \begin{tabular}{lccc}
    \hline
       Method  & AUC-ROC & Accuracy & F1-score \\ \hline
       Multi-resolution & 0.63 & 0.28 & 0.26 \\
       ResNet-50 & 0.58 & 0.29 & 0.25 \\
       AlSubaie \etal \cite{alsubaie_growth_2023}  & 0.57 & 0.28 & 0.26 \\
       Ding \etal \cite{Rina_tailoritingPretext}  & 0.59 & 0.25 & 0.18 \\
       \textbf{ViT-16} \cite{ViT} & \textbf{0.83} & \textbf{0.54} & \textbf{0.50} \\ \hline
       \textbf{\textit{CellOMaps}}  & \textbf{0.81} & \textbf{0.48} & \textbf{0.40} \\ \hline       
    \end{tabular}
    \label{tab:NLST_extVal}
\end{table}

\begin{table}
    \caption{External validation results for growth pattern classification on the ANORAK dataset.}
    \centering
    \vspace{0.5em}
    \begin{tabular}{lccc}
    \hline
       Method  & AUC-ROC & Accuracy & F1-score \\ \hline
       ResNet-50 & 0.64 & 0.36 & 0.20 \\
       Ding \etal \cite{Rina_tailoritingPretext}  & 0.52 & 0.25 & 0.14 \\
       \textbf{ViT-16} \cite{ViT} & \textbf{0.91} & \textbf{0.79} & \textbf{0.65} \\
        \hline
       \textbf{\textit{CellOMaps}}  & \textbf{0.86} & \textbf{0.78} & \textbf{0.61} \\ \hline       
    \end{tabular}
    \label{tab:ANORAK_extVal}
\end{table}

\subsection{Qualitative results}
\label{subsec45}
\begin{figure*}
\centering
\includegraphics[width=\textwidth]{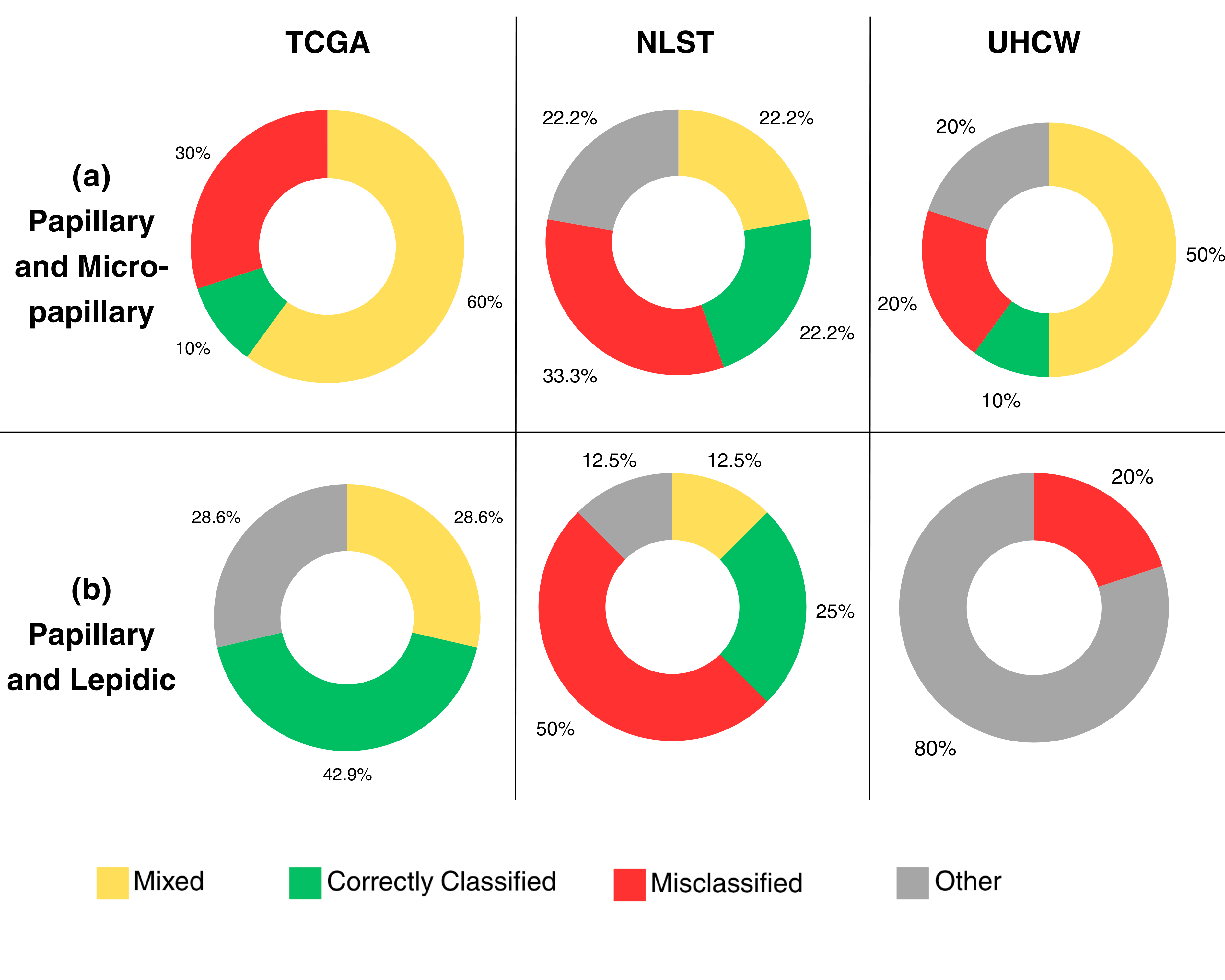}
\caption{Pathologists assessment of the misclassified tiles. Green indicates agreement with the predicted label, red indicate agreement with the ground truth, yellow indicates the appearance of mixed patterns in the tile, and gray meaning the tile expresses something other than the predicted label or ground truth. }
\label{question_summary}
\end{figure*}

\begin{figure*}
\centering
\includegraphics[width=0.85\textwidth]{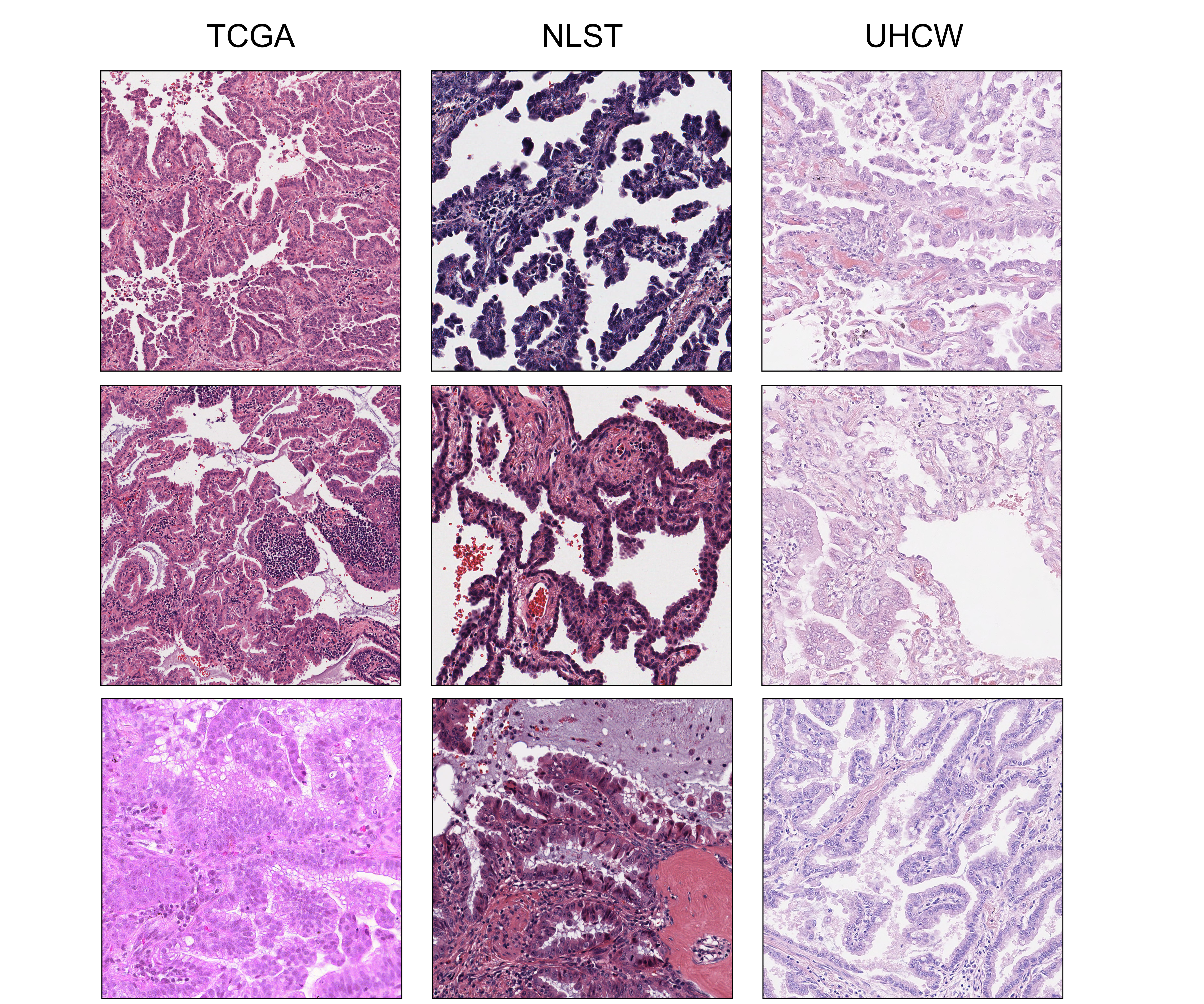}
\caption{Sample of misclassified tiles from TCGA (left column), NLST (middle column) and UHCW (right column). The rows depict confusion between different patterns: papillary and micropapillary (first row), papilllary and lepidic (second row), and papillary and acinar (third row). }
\label{sample_misclassified }
\end{figure*}

In our study, we also conduct some qualitative analysis to better understand the weakness of the proposed model. A subset of 60 misclassified tiles from all datasets were randomly selected, focusing on the most challenging classes, namely the misclassification of the papillary pattern as micropapillary or lepidic and vice versa. These tiles were given to two pathologists as presented to the model, with no further context, and they were each asked independently if the tile expressed the predicted pattern (correctly classified), the ground truth (missclassified), a mixture of those two patterns, or it expresses another pattern or morphology. To prevent bias, only the pattern names were given to the pathologist without specifying what is the original ground truth. Figure \ref{question_summary} shows the results of such assessment for the papillary and micropapillary pattern (a) and papillary and lepidic pattern (b).
We observed that the majority of the misclassified tiles actually express a mixture of patterns, both the ground truth and the predicted pattern. However, a few of the misclassified micropapillay tiles expressed a rare representation of the pattern that was not present in the training sample. A sample of the misclassified tiles are presented in Figure \ref{sample_misclassified }, where the lack of pattern purity is clearly evident.  

\subsection{Tumor Mutation Burden (TMB) prediction}
To demonstrate the effectiveness of the proposed model, we conducted preliminary experiments to provide indicative results and insights into the relationship between TMB levels and growth patterns. Several studies have investigated the differences in genomic profiles \cite{TMBRelationgenom} and TMB levels of LUAD tumors, based on their predominant growth patterns \cite{TMBRelation1, TMBRelation2}. High TMB is defined as having 10 or more mutations per megabase (mut/Mb), a threshold for which the U.S. Food and Drug Administration (FDA) has approved the use of pembrolizumab \cite{FDAapprove}. These studies show that high-grade tumors, namely solid and micropapillary predominant, are linked with high TMB levels \cite{TMBRelation1}. Conversely, low grade tumors, especially lepidic predominant tumors, are associated to low TMB levels and have more favorable outcomes \cite{lowTMBLep}. 
To visualize the distinction between high and low TMB tumors, we overlay the model's predictions on WSIs with varying TMB levels. As depicted in Figure \ref{model} (c), we observe that the growth patterns in patients with low TMB tend to be clustered and have a clear transition between one another (well differentiated), compared to more scattered and intermixed patterns (poorly differentiated) in patients with high TMB. Moreover, we observe the presence and spread of high grade patterns (micropapillary and solid) in patients with high TMB compared to patients with low TMB.

Furthermore, we use our proposed pipeline to predict all growth patterns present in the entire non-mucinous TCGA-LUAD cohort, consisting of 372 WSIs. We then represent each WSI by a 6-feature vector representing the percentage of each pattern in the tumor along with the percentage of normal tissue. These feature vectors were than fed to a simple Multi-Layer Perception network (MLP) with a single hidden layer to classify the patient as having low or high TMB. Using 5-fold cross validation an average accuracy of 0.66 ± 0.08 was achieved. Compared to an average accuracy of 0.63 ± 0.05 when applying IDaRS \cite{idars} on the same splits. 

We also investigate incorporating the spatial placement and adjacency of the patterns in a WSI, and study its impact on TMB prediction. In this experiment we proposed CellOMaps GP-GNN, a Graph Neural Network (GNN) were each node is a tile represented by a 6-feature vector consisting of the predicted probability of each pattern, and normal, output by our proposed classification model. The edges of the graph connect each node to its four nearest neighbors, forming a grid like network mimicking the tile placements in the WSI. This slightly improves the average accuracy to 0.67 ± 0.04. 

Finally to ensure that this improvement is an added value of the pattern's spatial information and not caused by the use of GNNs. We used Slidegraph+ \cite{slidegraph+}, where we represented the node by the 15 cell shape morphology features proved as descriptors in the paper along with deep features from ResNet-50. The remaining GNN architecture was maintained. This approach achieved an average accuracy of 0.41 ± 0.02, falling about 20\% behind in accuracy. Results are summarized in Table \ref{tab:TMB}

\begin{table}
    \caption{TMB prediction using 5-fold cross validation on the TCGA-LUAD cohort}
    \centering 
    \vspace{0.5em}
    \begin{tabular}{lccc}
        \hline 
        Method & AUC-ROC & Accuracy \\ \hline
        IDaRS \cite{idars} & 0.62 ± 0.05 & 0.63 ± 0.05 \\
        Slidegraph+ \cite{slidegraph+} &  0.66 ± 0.03 & 0.42 ± 0.03  \\
        CellOMaps GP-MLP  & 0.64 ± 0.07 & 0.66 ± 0.08  \\
        \textbf{CellOMaps GP-GNN}  & \textbf{0.67 ± 0.04} & \textbf{0.67 ± 0.4} \\
         \hline
    \end{tabular}
    \label{tab:TMB}
\end{table}

We will perform further investigation to better understand the role of growth patterns in predicting TMB and other important factors such as prognosis in LUAD in future work.

\subsection{Ablation Studies}
To optimize our proposed framework and asses the various components' added value we conduct ablation studies using the TCGA-LUAD dataset detailed in section \ref{secTCGA}. Each experiment is done using  5-fold cross validation adopting patient-level splits. 

\subsubsection{\textit{CellOMaps} structure}
To systematically evaluate the effectiveness and sufficiency of our proposed \textit{CellOMaps} representation compared to the standard H\&E, we employed ResNet-50 trained with focal loss, with varying input representations, all maintaining the same field of view. We compared the model's performance with the following inputs: an H\&E tile at 0.5 mpp, an H\&E tile at 2 mpp, and our proposed 3-channel \textit{CellOMaps}.

Additionally, we modified the first layer of the pretrained ResNet-50 to accept a 6-channel input image, where we stacked the H\&E tile at 2 mpp with its corresponding \textit{CellOMaps}. This modification aimed to determine whether the model could leverage both morphological features and abstract cellular arrangements.

Furthermore, we adjusted the first layer of the pretrained ResNet-50 to accept a single-channel grayscale image representing all cells in one mask without type separation. This arrangement was crucial to assess the importance of stratifying by cell types.

The results, listed in Table \ref{tab:input_abilation}, demonstrate the superior performance of the proposed \textit{CellOMaps} representation. These findings indicate that the cellular arrangement of different cell types alone is sufficient to distinguish between various growth patterns, and the cell type information is important as without those the performance drops to be equivalent to random guessing.

\begin{table*}
    \caption{Effect of varying input representation on the performance of growth pattern classification on the internal unseen test set. }
    \centering
    \vspace{0.5em}
    \begin{tabular}{lccc}
        \hline 
        Input & AUC-ROC & Accuracy & F1-score \\ \hline
        H\&E (0.5 mpp) & 0.56 ± 0.04 & 0.23 ± 0.03 & 0.18 ± 0.02\\
        H\&E (2 mpp) &  0.59 ± 0.04 & 0.24 ± 0.07 & 0.18 ± 0.02 \\
        \textbf{3 channel \textit{CellOMaps}} (2 mpp) & \textbf{0.92 ± 0.03} & \textbf{0.81 ± 0.02}  & \textbf{0.7 ± 0.04} \\
        H\&E + \textit{CellOMaps} (2 mpp) & 0.84 ± 0.03 & 0.62 ± 0.1 & 0.51 ± 0.03 \\
        single channel \textit{CellOMaps} (2 mpp) & 0.49 ± 0.01 & 0.3 ± 0.07 & 0.08 ± 0.01  \\
         \hline
    \end{tabular}
    \label{tab:input_abilation}
\end{table*}

\subsubsection{Network architecture}
The choice of backbone network architecture is paramount to achieving optimal performance. As discussed in section \ref{sec2} the majority of the work in the literature utilize the power of pre-trained CNNs for the classification of LUAD growth patterns, with ResNet-50 being the most popular in such application. Moreover, the recent introduction of the vision transformer, that rely on the self attention mechanism,  have boosted the performance and prediction accuracy on natural images \cite{ViT}. In this section, we compare the results of using different backbone architectures for classifying \textit{CellOMaps} into the different growth patterns or as normal. We experiment using ResNet-18 (pretained on ImageNet), ResNet-50 (pretained on ImageNet), ResNet-50 with two extra self attention layers (SA) before the final fully connected layer, and the vision transformer (ViT-16, pretained on ImageNet). AUC-ROC, accuracy, and F1-scores reported in Table \ref{tab:backbone_abilaation} show that ResNet-50 provides sufficient performance without the need of more complex and computationally expensive models. 

\begin{table}[h!]
    \caption{Effect of using different classification backbone on the performance of growth pattern classification in the proposed pipeline on the internal unseen test set. }
    \centering
    \vspace{0.5em}
    \resizebox{\columnwidth}{!}{
    \begin{tabular}{lccc}
        \hline 
        Backbone & AUC-ROC & Accuracy & F1-score \\ \hline
        ResNet-18 & 0.91 ± 0.01  & 0.73 ± 0.05 & 0.62 ± 0.04 \\
        \textbf{ResNet-50} & \textbf{0.92 ± 0.03} & \textbf{0.81 ± 0.02}  & \textbf{0.7 ± 0.04} \\
        ResNet-50 + SA & 0.91 ± 0.02 & 0.79 ± 0.04 & 0.68 ± 0.02 \\
        ViT-16 & 0.9 ± 0.03 & 0.65 ± 0.04 & 0.47 ± 0.06 \\
         \hline
    \end{tabular}}
    \label{tab:backbone_abilaation}
\end{table}

\subsubsection{Loss function}
The selection of an appropriate loss function is a critical component in the design of deep learning models, as it directly influences the convergence behavior and overall performance. In this section, we systematically examine the impact of the loss function on the proposed model's performance. We examine the classic cross entropy, weighted cross entropy (with static class weights based on the number of samples belonging to the class), and focal loss (which employs a dynamic weight scaling). The later two are to address the class imbalance problem present in growth patterns. Results listed in Table \ref{tab:loss_functions}  show that focal loss improves the per class F1-scores compared to cross entropy and weighted cross entropy. In our experiments, we empirically found that setting $\gamma = 0.7$ worked best for our specific problem.

\begin{table*}[h!]
    \caption{Average and standard deviation of F1-scores per class on the internal unseen test set for different loss functions. (CE): cross entropy, (WCE): Weighted cross entropy, (FL): Focal loss.}
    \centering
    \vspace{0.5em}
    \resizebox{\textwidth}{!}{%
    \begin{tabular}{@{}lcccccccc@{}}
    \hline
        loss function & solid & Acinar & Papillary & Micropapilaary & lepidic & Normal \\ \hline
         CE & \textbf{0.95 ± 0.04} & \textbf{0.9 ± 0.05} & 0.45 ± 0.22 & 0.58 ± 0.17 & 0.76 ± 0.11 & 0.98 ± 0.03 \\
         WCE & 0.89 ± 0.12 & 0.84 ± 0.12 & \textbf{0.71 ± 0.23} & 0.64 ± 0.17 & 0.77 ± 0.16 & 0.98 ± 0.01 \\
         \textbf{FL} & 0.93 ± 0.08 & 0.84 ± 0.08 & 0.59 ± 0.21 & \textbf{0.67 ± 0.11} & \textbf{0.87 ± 0.05} & \textbf{0.99 ± 0} \\
        \hline
    \end{tabular}
    }
    \label{tab:loss_functions}
\end{table*}

\subsection{Image size}
\textit{CellOMaps} require only 3 bits to represent a pixel (1 bit/channel), compared to 24 bits needed to represent a RGB-image pixel (8 bits/channel).
To demonstrate the compactness and efficiency of our proposed \textit{CellOMaps} representation for WSIs, we quantified its compactness by calculating the Shannon entropy \cite{shannon1948mathematical} of 448×448 tiles at 1 mpp. This metric provides a rigorous assessment of the information density within an image.

\begin{figure*}
\centering
\includegraphics[width=\textwidth, height=0.85\textheight,keepaspectratio]{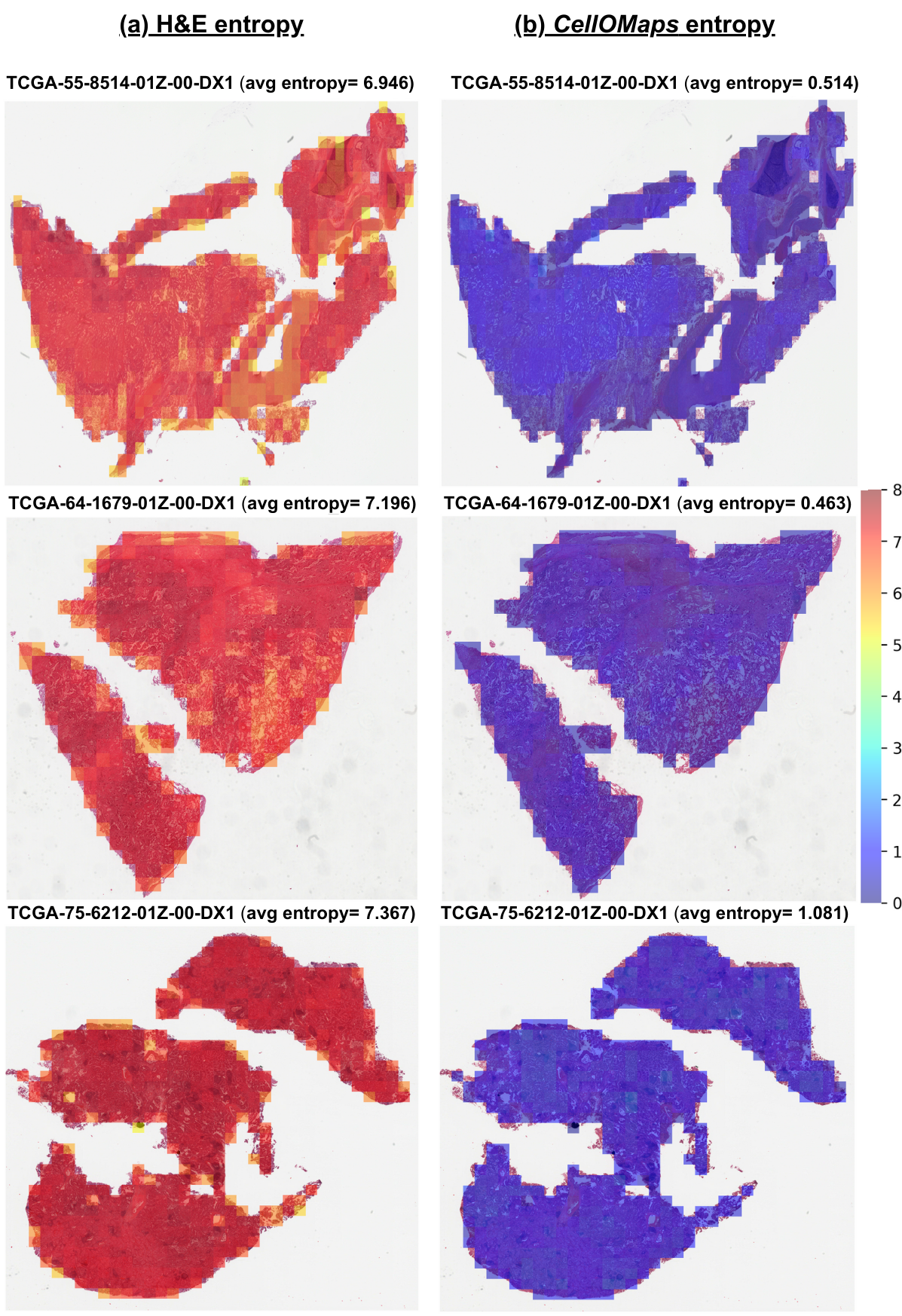}
\caption{Heatmaps of the Shannon entropy for sample slides. (a) H\&E representation, (b)\textit{CellOMaps} representation.}
\label{entropy}
\end{figure*}

We compare the average Shannon entropy of all H\&E stained tiles in our TCGA-LUAD dataset to their corresponding \textit{CellOMaps}, This quantification highlights the reduction in information achieved by our method reducing the average Shannon entropy of a tile from 7.20 bits in the H\&E images, to 0.74 bits in the proposed \textit{CellOMaps}. Figure \ref{entropy} illustrates such reduction on sample slides from TCGA-LUAD showing heatmaps for the Shannon entropy of both H\&E and \textit{CellOMaps}.     

\section {Discussion}

Identifying growth patterns in lung adenocarcinoma tumors is crucial for assessing tumor grade and determining appropriate treatment. However, these patterns are highly heterogeneous, leading to significant inter- and intra-observer variability. Currently, pathologists typically report only the dominant pattern, sometimes listing other significantly represented patterns. We believe that identifying all patterns in a whole slide image (WSI) along with their spatial relationships could provide a better understanding of tumor behavior leading to a better diagnosis.

Recent studies have attempted to use machine learning to predict growth patterns, predominantly focusing on predicting a single dominant pattern for a WSI. Although some studies have developed tile-based classifiers, they often lack proper evaluation since performance metrics are reported on unseen tiles rather than unseen WSIs.

To address these clinical challenges and fill the current gap in the literature, we propose a novel pipeline for growth pattern classification. Our proposed approach substitutes the H\&E stained image with a simpler and lighter format called \textit{CellOMaps}, which maintain cellular information and spatial arrangement. Moreover, we conducted an extensive evaluation of the proposed model and performed a fair comparison with state-of-the-art tile-based growth pattern classifiers. Our results demonstrate that with appropriate data splitting, most models proposed in the literature fail, achieving average F1-scores around 0.2 on internal cross-validation using the TCGA-LUAD dataset. In contrast, our proposed pipeline achieves an outstanding F1-score of 0.7. Validation results on external datasets yield similar outcomes, reinforcing our findings.

Our preliminary findings suggest that the model's output can be used to effectively distinguishes between high and low TMB tumors, highlighting the association between high TMB levels and the presence of aggressive growth patterns such as micropapillary and solid. These results align with existing literature. Moreover, we show that exploiting the full spatial information of these patterns can enhance TMB prediction. However, further research with advanced methods is essential to enhance these results and validate the significance of such observations.

The proposed approach demonstrates that cell types and their spatial relationships are sufficient to distinguish between different growth patterns. However, this imposes a high dependency on the nuclei detection and classification algorithm used, as its precision significantly impacts the model's final classification.

Another limitation of the current study, and other similar studies, is the lack of regional annotations, which are costly to obtain as they require a minimum of two trained pathologists to annotate the same regions. To facilitate further research,our annotations for the TCGA-LUAD slides have been made publicly available.

\section{Conclusions}
In this paper, we introduced \textit{CellOMaps}, a novel compressed representation of the H\&E WSI, that effectively captures cellular composition. We proposed a classification model that utilizes these \textit{CellOMaps} to classify LUAD growth patterns. Through extensive and proper evaluation (using patient-level splitting), we show that the proposed pipeline achieves state-of-the-art performance compared to current growth pattern classification models. Moreover, we showcase the generalizability of the proposed model via external validation. Finally, we highlight the impact of our model predictions, and its potential use for TMB stratification. In light of the results presented here, in the future we plan to extend our analysis to predict prognosis, single gene mutations, and patient outcomes such as survival ultimately enhancing our understanding of lung adenocarcinoma.

\section*{Data and Code availability}
We provide the code and the annotation masks for TCGA-LUAD dataset in the following repository (\href{URL}{https://github.com/Arwa-AlRubaian/CellOMaps}). Interested reader may use the code and masks to replicate the the results presented in this paper.  


\section*{Acknowledgment}
AR is supported by the Saudi Cultural Bureau in London, UK. SEAR reports financial support from the MRC (MR/X011585/1) and the BigPicture project, funded by the European Commission. NR report financial support from GlaxoSmithKline, United Kingdom, outside of the submitted work. NR is CEO and CSO of Histofy Ltd.
The authors would like to thank Ruiwen Ding and William Hsu from  University of California, Los Angeles for sharing NLST annotations and their model checkpoints.   

\appendix
\section{Weak validation approach, when adopting a tile-based data splitting}
\label{app1}
We present the results of evaluating the proposed approach and comparing it to state-of-the-art methods using the weak validation approach widely adopted in the literature which involves employing a tile-based data splitting approach. We apply 5-fold cross validation at the tile level on a subset of 18 WSIs from TCGA-LUAD. We include the results of this weak validation solely for comparison purposes. We suggest strong validation should be used for testing robustness of the algorithms.

\begin{table}
    \caption{Quantitative comparison of the performance of the proposed method along with other state-of-the-art and baseline methods for growth pattern classification using tile-based data splitting}
    \centering
    \vspace{0.5em}

    \begin{tabular}{lccc}
        \hline 
        Method & AUC-ROC & Accuracy & F1-score \\ \hline
        Multi-resolution  & 1.0 ± 0.0&  0.96 ± 0.01&  0.97 ± 0.01 \\
        ResNet-50 & 1.0 ± 0.0&  0.96 ± 0.01&  0.48 ± 0.0 \\
        AlSubaie \etal \cite{alsubaie_growth_2023} & 0.94 ± 0.01&  0.66 ± 0.03&  0.68 ± 0.04 \\
        \hline
        \textit{CellOMaps} (CE) & 0.97 ± 0.01&  0.81 ± 0.03&  0.89 ± 0.02 \\
       \hline
    \end{tabular}
    
    \label{tab:internal_results}
\end{table}.

\section{Extra ablation Studies}
\label{app2}
\subsection{CellOMaps Cell Types}
The subset of cells to include in the proposed \textit{CellOMaps} representation is a critical choice, as it forms the network learning material. HoVer-Net classifies a nuclei to one of the five types: Neoplastic epithelial, non-neoplastic epithelial, connective tissue, inflammatory, and necrosis. In this section, we study the affect of excluding and including different cell types in our proposed \textit{CellOMaps}. We include Neoplastic epithelial cells in all formations (as they represent the tumor cells), and exclude necrosis due to its irrelevance. 
We construct the \textit{CellOMaps} by placing each cell type in a channel and use an appropriate adaptation of ResNet-50 to classify the growth patterns. 
Table \ref{tab:cellType_abilation} reports the average and standard deviation of five distinct runs on internal unseen test sets from TCGA-LUAD. Results show that the most representative \textit{CellOMaps} is the one composed of neoplastic epithelial, non-neoplastic epithelial, and connective cells. This can be justified by the definition of the growth patterns \cite{whoClassification}, where we observe that the inclusion of connective cells enhances the prediction of the papillary and acinar patterns, thus improving the overall F1-score. While the placement of inflammatory cells do not define any of the patterns and thus their inclusion only confuses the model.

\begin{table*}
    \caption{Effect of including and excluding different cell types in the formation of the \textit{CellOMaps} on the performance of growth pattern classification on the internal unseen test set. Tepith (Neoplastic epithelial), Nepith (Non-Neoplastic epithelial), Conn (connective), Inf (inflamatory). }
    \centering
    \vspace{0.5em}
    \begin{tabular}{lccc}
        \hline 
        Cell types included & AUC-ROC & Accuracy & F1-score \\ \hline
        Tepith  & 0.85 ± 0.11 &  0.69 ± 0.05 & 0.61 ± 0.03 \\
        Tepith + Nepith  & 0.66 ± 0.06 & 0.33 ± 0.08 & 0.22 ± 0.03 \\
        Tepith + Conn & 0.84 ± 0.01 & 0.56 ± 0.11 & 0.51 ± 0.04 \\    
        Tepith + Inf & 0.84 ± 0.03 & 0.55 ± 0.09  & 0.49 ± 0.03 \\     
        Tepith + Inf + Nepith  & 0.79 ± 0.12 & 0.63 ± 0.07 & 0.58 ± 0.05  \\
        Tepith + Inf + Conn  & 0.88 ± 0.02 & 0.72 ± 0.08 & 0.59 ± 0.03  \\
        Tepith + Inf + Nepith + Conn & 0.65 ± 0.06  & 0.4 ± 0.08 & 0.29 ± 0.09  \\
        \textbf{Tepith + Nepith + Conn} &  \textbf{0.92 ± 0.03} & \textbf{0.81 ± 0.02}  & \textbf{0.7 ± 0.04}  \\
        
         \hline
    \end{tabular}
    \label{tab:cellType_abilation}
\end{table*}

\subsubsection{Tile size}
\begin{figure*}[h]
\centering
\includegraphics[height=0.4\textheight]{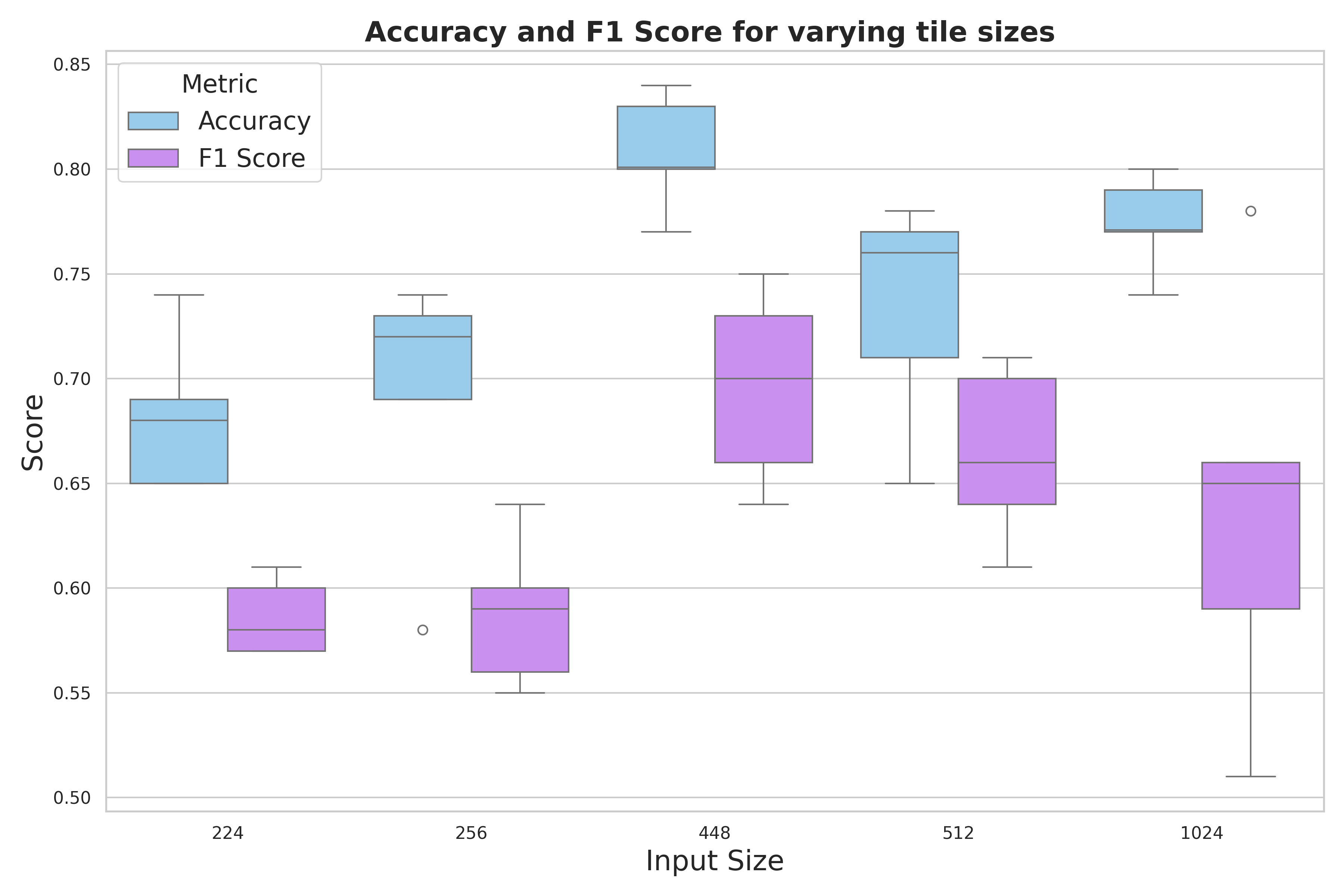}
\caption{Accuracy and F1-score for varying \textit{CellOMaps} tile sizes.}
\label{tileSize_abliation}
\end{figure*}

The \textit{CellOMaps} provide a compressed version of the WSI that conserves the full cellular composition; as it captures the detailed cellular arrangement at 0.5 mpp and compresses it by a factor of 4 without information loss. This allows us to capture more context in a single tile compared to its H\&E equivalent. However, as our growth pattern labels are regional, and a single WSI usually contain more than one pattern we needed to find the "perfect" tile size that captures enough context to identify the pattern without spanning more than a single pattern. We experiment with different tile sizes  starting at 224×224 up to 1024×1024. Results illustrated in Figure \ref{tileSize_abliation} show that a tile size of 448×448 is the best fit for our problem. 

\section*{Declaration of generative AI and AI-assisted technologies in the writing process}

During the preparation of this work the author(s) used ChatGPT in order to enhance readability and sentence structure. After using this tool/service, the author(s) reviewed and edited the content as needed and take(s) full responsibility for the content of the publication.

\bibliographystyle{unsrt.bst}
\bibliography{references.bib}

\end{document}